# Faster estimation of the correlation fractal dimension using box-counting

Christos Attikos
*Accenture S.A.,
Management Consulting Division
1, Arkadias Str., 14564 Kifisia, Greece*
christos.attikos@accenture.com

Michael Doumpos
*Department of Production Engineering and
Management, Technical University of Crete
University Campus, 73100 Chania, Greece*
mdoumpos@dpem.tuc.gr

## Abstract

*Fractal dimension is widely adopted in spatial databases and data mining, among others as a measure of dataset skewness. State-of-the-art algorithms for estimating the fractal dimension exhibit linear runtime complexity whether based on box-counting or approximation schemes. In this paper, we revisit a correlation fractal dimension estimation algorithm that redundantly rescans the dataset and, extending that work, we propose another linear, yet faster and as accurate method, which completes in a single pass.*

## 1. Introduction

Considering datasets as self-similar at different scales, thus fractals [1], facilitates their characterization with one single measure, their intrinsic or fractal dimension. Database research has prolifically employed the concept of fractal dimension in various areas including selectivity estimation [2], dimensionality reduction [3], outlier detection [4], and clustering [5].

One definition of the fractal dimension is the correlation fractal dimension $D_2$ [2], [6]:

$$D_2 \equiv \frac{\partial \log \sum_i C_i^2}{\partial \log(r)}, r \in [r_1, r_2]$$

where $C_i$ is the occupancy of the $i$-th cell when the address space is embedded into an $E$-dimensional grid with cells of side $r$.

Algorithms for estimating $D_2$ range in time and space complexity. Table 1 shows the most prominent ones and summarizes their attributes.

**Table 1. Algorithms for estimating the correlation fractal dimension**

| Work | Method used | Time complexity[1] | Space complexity |
|---|---|---|---|
| [7] | Pair-counting | $O(N^2)$ | $O(N)$ |
| [2] | Box-counting | $O(N\log N)$ | $O(N)$ |
| [3] | Box-counting | $O(EN|R|)$ | $O(N)$ |
| [8] | Approximation | $O(EN|R|/s_1 s_2)$ | $O(1)$ |

Among the aforementioned efforts, the linear box-counting Fractal Dimension algorithm (FD) [3] institutes the state-of-the-art from a runtime perspective. Extending that work, we propose the Faster Fractal Dimension (FFD) algorithm, which scans the dataset just once, performs independently of the number of different radii taken into account, and is by definition as accurate as its predecessor.

The paper proceeds as follows. We summarize the original FD algorithm in the next section. Section 3 introduces the novel FFD algorithm explaining key features. Section 4 discusses some experimental evaluation demonstrating the effectiveness of the new algorithm. Finally, Section 5 summarizes this work.

---

[1] In the mentioned papers, |R| values ranged from 10 to 31 while $s_1$ and $s_2$ where equal to 30 and 5 respectively.

## 2. Fractal Dimension algorithm (FD)

Figure 1 shows the FD algorithm [3]. Cells are uniquely identified using a row-wise space filling curve. FD imposes grids of different resolutions over the dataset starting from the coarsest one (Figure 2), counts the cells' occupancies and creates the log($r$)-log($S(r)$) plot.

---
**Algorithm FD**(normalized dataset $A$)
1. For each grid of size $r = 1/2^j, j = 1, 2, ..., |R|$
2.    For each point of the dataset
3.       Decide which cell it falls in
4.       Increment the occupancy counter $C_i$
5.    Compute the sum of squared occupancies $S(r) = \sum_i (C_i)^2$
6. Print the values of log($r$) and log($S(r)$) generating a plot.
7. Return the slope of the linear part of the plot as $D_2$ of dataset $A$.

**End FD**

---
**Figure 1. Fractal Dimension algorithm [3]**

The runtime complexity of the FD algorithm is O($EN|R|$), where $N$ is the number of objects in the dataset, $E$ is the number of attributes of the dataset and $|R| = |\{r\}|$ is the number of radii used (the number of points in the resulting box-count plot).

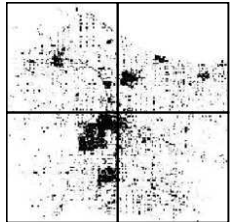

**(a)** $r = 1/2^1$

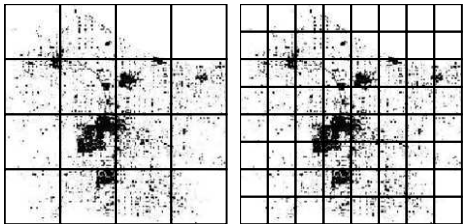

**(b)** $r = 1/2^2$     **(c)** $r = 1/2^3$

**Figure 2: Reconsidering the dataset for every grid resolution**

Re-examining the dataset for each grid resolution appears counter-intuitive and suboptimal. As an alternative, we should be able to use previously calculated data for subsequent grids. This is the rationale behind our approach.

## 3. Faster Fractal Dimension algorithm (FFD)

The proposed improved algorithm is illustrated in Figure 3. Here, the sequence of grid granularities is reversed; hence, we start by filling in the occupancies of the finest grid and proceed consecutively with coarser-resolution grids.

We exploit the fact that each cell of the next-step grid (i.e. coarser) overlaps $2^E$ cells of the previous-step one (i.e. finer). Thus, instead of reconsidering the entire dataset, we sum up the occupancies of cells that belong at the same "parent" cell of the next-step grid (see example in Figure 4).

---
**Algorithm FFD**(normalized dataset $A$)
1. $j = |R|$   //finest grid resolution
2. For each point of the dataset
3.    Decide which cell in $Grid_j$ it falls in
4.    Increment the occupancy counter $C_{j,i}$
5. Compute the sum of squared occupancies $S(r) = \sum_i (C_{j,i})^2$
6. For each $Grid_j$ of size $r = 1/2^j, j = |R|-1, ..., 1$
7.    For each occupied cell in $Grid_{j+1}$
8.       Decide which cell in $Grid_j$ it falls in
9.       Increment the occupancy counter $C_{j,i}$ by $C_{j+1,i}$
10.    Compute the sum of squared occupancies $S(r) = \sum_i (C_{j,i})^2$
11. Print the values of log($r$) and log($S(r)$) generating a plot.
12. Return the slope of the linear part of the plot as $D_2$ of dataset $A$.

**End FFD**

---
**Figure 3. Faster Fractal Dimension algorithm**

In terms of execution time the gain is two-fold: The dataset is scanned only once and the search space is decreased by $1/2^E$ at each step. Furthermore, less memory is required, since FFD considers only two successive grids per step while FD uses the entire list of grids.

FFD's runtime complexity is upper-bounded by $EN + EN\sum_j 1/2^{E(j-1)}$ for $j = 1$ to $|R|$, thus it is O($cEN$), with

$c = 1 + 2^E/(2^E-1)$. The resulting algorithm scales independently of $|R|$.

For example with 2D data, $c$ equals to 2.333 while $|R|$ could range from 10 to 30. For higher dimensionalities, the runtime difference between FFD and FD increases ($c$ converges to 2). To the best of our knowledge, this is the fastest fractal dimension estimation algorithm in the database literature.

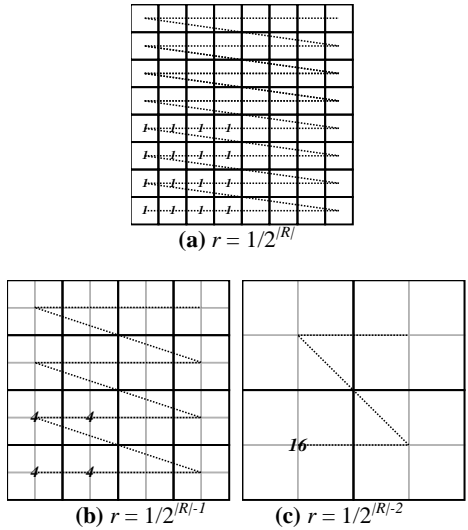

**Figure 4. Adding the occupancies of overlapping cells between consecutive grids**

## 4. Experiments

Both algorithms were implemented with Microsoft Visual Studio .NET [9] and experiments were executed using a PC with a Pentium IV 1.8GHz processor and 320 MB of RAM under Windows XP.

Real datasets along with synthetic ones with known theoretical $D_2$ values where used and as anticipated both algorithms provided identical results. Figure 6 shows an example box-count plot. The slope of the linear part of the plot is the estimated $D_2$ of the Sierpinski dataset (Figure 5).

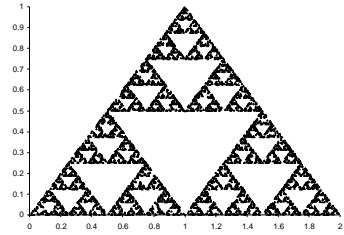

**Figure 5. The Sierpinski dataset; Theoretical $D_2$ = 1.585**

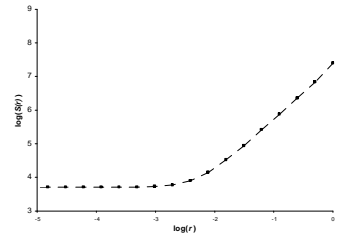

**Figure 6. Resulting box-count plot for the Sierpinski dataset; Estimated $D_2$ = 1.588**

All datasets where normalized to the unit space and radius values adhered to the progression $\alpha_j = 1/2^{|R|-j}$ for $j = 1$ to $|R|$ (with $|R| = 10$). Grids where implemented as hash tables and cells where added on demand, as occupancies occurred.

Figure 7 illustrates the superior runtime performance of the FFD algorithm. Two-dimensional synthetic datasets comprised of points uniformly distributed in the unit space were used. All data was loaded in main memory so there is no I/O overhead.

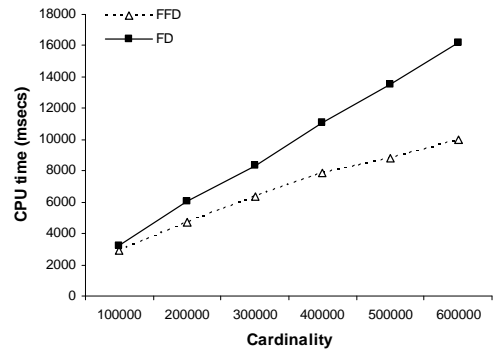

**Figure 7. Running time versus dataset size**

## 5. Conclusions

Our focus on fractal dimension is due to its ability to summarize the whole dataset in one value, i.e. its intrinsic dimension. Database research has successfully applied this notion in an effort to tackle various problems and several methods have been developed for estimating the fractal dimension.

In this work, we describe FFD, the fastest -to the best of our knowledge- algorithm for estimating the fractal dimension of a dataset. Compared to FD, an earlier algorithm, FFD scans the dataset only once while reducing the search space at each step. Experiments using synthetic data verify the superior performance of FFD and support its use in high-dimensional databases.

## 6. References

[1] Mandelbrot, B., *The Fractal Geometry of Nature*, W.H. Freeman, New York, 1977.

[2] A. Belussi, and C. Faloutsos, "Estimating the selectivity of spatial queries using the 'correlation' fractal dimension", *Proceedings of VLDB conference*, 1995.

[3] C. Traina Jr., A.J.M. Traina, L. Wu, and C. Faloutsos, "Fast feature selection using the fractal dimension", *Proceedings of Brazilian Symposium on Databases*, 2000.

[4] S. Papadimitriou, H. Kitagawa, P.B. Gibbons, and C. Faloutsos, "LOCI: fast outlier detection using the local correlation integral", *Proceedings of ICDE conference*, 2003.

[5] D. Barbará, and P. Chen, "Using the fractal dimension to cluster datasets", *Proceedings of ACM SIGKDD conference*, 2000.

[6] P. Grassberger, and I. Procaccia, "Measuring the Strangeness of Strange Attractors", *Physica D: Nonlinear Phenomena*, vol. 9, nos. 1-2, pp. 189-208, 1983.

[7] P. Grassberger, "An optimized box-assisted algorithm for fractal dimensions", *Physics Letters A*, vol. 148, nos. 1-2, pp. 63-68, 1990.

[8] A. Wong, L. Wu, P.B. Gibbons, and C. Faloutsos, "Fast estimation of fractal dimension and correlation integral on stream data", *Information Processing Letters*, vol. 93, pp. 91-97, 2005.

[9] Microsoft Corporation, *Microsoft Visual Studio .NET*, Available at http://msdn.microsoft.com/vstudio/ (accessed November 20, 2008).